# Aerodynamic Prediction of High-Lift Configuration Using $k - \overline{v^2} - \omega$ Turbulence Model

Shaoguang Zhang, [1,*] Haoran Li, [1,†] Yufei Zhang, [1,‡] and Haixin Chen [1,§]
*(1. Tsinghua University, Beijing, 100084, People's Republic of China)*

**The aerodynamic performance of the high-lift configuration greatly influences the safety and economy of commercial aircraft. Accurately predicting the aerodynamic performance of the high-lift configuration, especially the stall behavior, is important for aircraft design. However, the complex flow phenomena of high-lift configurations pose substantial difficulties to current turbulence models. In this paper, a three-equation $k - \overline{v^2} - \omega$ turbulence model for the Reynolds-averaged Navier-Stokes equations is used to compute the stall behavior of high-lift configurations. A separated shear layer fixed function is implemented in the turbulence model to better capture the nonequilibrium characteristics of turbulence. Different high-lift configurations, including the two-dimensional multielement NLR7301 and Omar airfoils and a complex full-configuration model (JAXA Standard Model), are numerically tested. The results indicate that the effect of the nonequilibrium characteristics of turbulence is significant in the free shear layer, which is key to accurately predicting the stall behavior of high-lift devices. The modified SPF $k - \overline{v^2} - \omega$ model is more accurate in predicting stall behavior than the Spalart-Allmaras, shear stress transport, and original $k - \overline{v^2} - \omega$ models for the full high-lift configuration. The relative errors in the predicted maximum lift coefficients are within 3% of the experimental data.**

## Nomenclature

*AOA* = angle of attack, deg

$C_D$ = drag coefficient

---

[*] Ph. D student, School of Aerospace Engineering, email: zsg@mail.tsinghua.edu.cn
[†] Ph. D student, School of Aerospace Engineering, email: lihr17@mails.tsinghua.edu.cn
[‡] Associate professor, School of Aerospace Engineering, senior member AIAA, email: zhangyufei@tsinghua.edu.cn (corresponding author)
[§] Professor, School of Aerospace Engineering, associate fellow AIAA, email: chenhaixin@tsinghua.edu.cn

| | | |
|---|---|---|
| $C_L$ | = | lift coefficient |
| $Cm$ | = | pitching moment coefficient |
| $C_p$ | = | pressure coefficient |
| c | = | airfoil chord length, m |
| $\rho$ | = | air density, kg/m$^3$ |
| $Re$ | = | Reynolds number |
| $S$ | = | shear rate, 1/s |
| $\mu$ | = | molecular viscosity, Pa.s |
| $\nu$ | = | kinematic molecular viscosity, $\mu/\rho$, $m^2/s$ |
| $\Omega$ | = | magnitude of vorticity, $1/s$ |

## I. Introduction

The high-lift configuration is widely used in commercial aircraft to compensate for the low velocity during take-off and landing [1]. The high-lift configuration usually has complex geometries, such as for slats, flaps, slat brackets, and flap fairings. These complex geometries induce complicated flow phenomena that pose challenges for numerical simulations. The complex flow structure includes wakes under adverse pressure gradients, wake/boundary-layer merging, streamline curvature flow, separated flow, possible unsteady flow, wing-tip vertical flow, shockwave/boundary-layer interaction, and laminar/turbulent transition regions on wing elements (slats, main wings or flaps) [2,9]. The high-lift configuration usually operates at large angles of attack. Hence, flow separation is likely to occur on the upper surfaces of wings and flaps [1]. For a typical high-lift configuration, the flow separation region usually starts from the wing root [17]. It is also possible that stalls are induced by the presence of flow separation behind the flap track fairings and slat brackets [3].

The AIAA High-Lift Prediction Workshop (HiLiftPW) series was organized to assess the state-of-the-art numerical prediction capabilities for commercial transport-type aircraft in landing and/or take-off configurations and to promote improvements to modeling and simulation capabilities [4]. The first HiLiftPW [4] in June 2010 focused on the effects of the grid type, grid density, solver, and turbulence model, and a standard model for a three-element swept wing with detailed experimental data was provided. Subsequently, in 2013, the DLR-F11 model in the landing configuration was used in the second HiLiftPW [5]. In addition to considering the effects of the grid density and turbulence model, the effects of the support brackets were also taken into account [5]. The third and most recent workshop, HiLiftPW-3, was held in 2017 [6]. The JAXA Standard Model (JSM) was adopted both without and with nacelles/pylons to study the effects of nacelles/pylons on high-lift flows. Many valuable conclusions have been obtained through these workshops, which have shown that the turbulence model has a great influence on the accuracy of predicting the aerodynamic performance (especially the stall behavior) of the high-lift configuration.

   Predicting the stall performance of the high-lift configuration requires a reliable turbulence model. A series of high-lift configurations have been experimentally tested to validate turbulence models [7-12]. Moreover, many high-fidelity unsteady CFD methods for high-lift flows have been promoted, such as the large-eddy simulation [13], detached-eddy simulation [14], and lattice Boltzmann method [17]. These methods can accurately predict the aerodynamic force coefficient of the high-lift configuration. However, the extremely high computational cost limits the application of these methods in the daily design process. Currently, the steady Reynolds-averaged Navier-Stokes (RANS) equations are still the main tools in the design of high-lift configurations. However, the present RANS models fail to predict the stall performance with large separation regions. Among the turbulence models, the two-equation shear stress transport (SST) model and the one-equation Spalart-Allmaras (SA) model are the most widely used for the design of high-lift configurations. Both models are fully turbulence models, which means that transition phenomena are not modeled. Yin [15] applied the SA model to evaluate a 30P30N multielement airfoil and found that the results match well with experimental data in the linear range but slightly overpredict the maximum lift coefficient at the stall angle of attack. Ashton [16] found that none of the SA, SST, and $k - \varepsilon$ models are able to capture the post-stall region in the case of the JSM. Marc et al. [18] took the laminar-turbulence transition into account and accurately predicted the stall behavior.

   Most of RANS models are calibrated in the equilibrium state of turbulence. The Richardson–Kolmogorov energy cascade has a constant dissipation coefficient ($C_\epsilon$) that describes the equilibrium energy transfer process [19]. The assumption that $C_\epsilon$ is a constant is the basis of modeling the turbulent viscosity coefficient [20]. Consequently, the flow phenomena that can be accurately described in most turbulence models are in an equilibrium state. However, many researchers [21,22,25] developed a new dissipation law that is not constant in experiments and numerical simulations, necessitating the study of nonequilibrium turbulence. In the classic study of free shear turbulence, the jet wake has a strong self-similarity [23]. Townsend [23] and George [24] thoroughly introduced the assumption-based inference that the equilibrium dissipation law is a constant. However, Nedić et al. [25] obtained a different similarity index from the classic conclusion based on the nonequilibrium dissipation law $C_\epsilon \sim Re_I Re_L$. The turbulence models calibrated by the equilibrium turbulence assumption can hardly be expected to provide an accurate description of nonequilibrium turbulence, which widely exists in free shear flows and near-stall flows. For example, Fang et al. [26] found that nonequilibrium turbulence is obvious in the regions of the corner separation, wake, and boundary layer of the compressor flow. Recently, Li et al. [27,28] developed a separating shear layer fixed (SPF) $k - \overline{v^2} - \omega$ model to simulate the stall behavior of iced airfoils; this three-equation model focused on the nonequilibrium characteristics of turbulence in separation regions. The effects of nonequilibrium characteristics in separated shear layer regions are reflected in the turbulence production $P_k$, which tends to be significantly larger than the turbulence dissipation $\varepsilon$ [29,30]. However, the influence of nonequilibrium characteristics on the accuracy of predicting the aerodynamic performance of high-lift configurations has scarcely been studied.

In this paper, a modified three-equation $k - \overline{v^2} - \omega$ model for calculating high-lift flows is developed. The nonequilibrium characteristics of turbulence are captured by adding a regional modification to the destruction term of the ω equation. Three high-lift configurations, including the NLR7301 and Omar multielement airfoils and a complex three-dimensional model (JSM), are used as the validation cases. Five RANS models, namely, the SA, SST, Wilcox06 $k - \omega$, original $k - \overline{v^2} - \omega$, and modified $k - \overline{v^2} - \omega$ models, are compared for different cases. Finally, the effect of the nonequilibrium characteristics of turbulence on predicting the aerodynamic performance of high-lift configurations is discussed.

## II. Numerical Method

### A. Numerical Solver

The aerodynamic analysis of the high-lift configuration in this work is carried out using the RANS solver CFL3D version 6.7 [31] with a structured grid. The spatial discretization for inviscid flux uses the monotone upstream-centered schemes for conservation laws (MUSCL) for the reconstruction and the Roe flux difference splitting method for the Riemann solver. The implicit approximate-factorization method is chosen for time advancement. Multigrid and mesh sequencing are provided to accelerate the convergence. All turbulence models are solved uncoupled from the Navier-Stokes equations using implicit approximate factorization.

### B. The SPF $k - \overline{v^2} - \omega$ model

The SPF k $-\overline{v^2}-\omega$ model has three equations, including the transport equations of the total fluctuation energy $k$ (both the full turbulence and the pretransition velocity fluctuations), $\overline{v^2}$ (the three-dimensional full turbulence fluctuations), and the specific dissipation rate $\omega$. The equations are listed below. In Eq. (2) and Eq. (3), the $R_{BP}$ and $R_{NAT}$ terms model the bypass and natural transition processes, respectively. The detailed formulation of each term is provided in reference [27].

$$\frac{\partial k}{\partial t}+u_j\frac{\partial k}{\partial x_j}=\frac{1}{\rho}P_k-min(\omega k,\omega\overline{v^2})-\frac{1}{\rho}D_k+\frac{1}{\rho}\frac{\partial}{\partial x_j}\left[\left(\mu+\frac{\rho\alpha_T}{\sigma_k}\right)\frac{\partial k}{\partial x_j}\right] \quad (1)$$

$$\frac{\partial\overline{v^2}}{\partial t}+u_j\frac{\partial\overline{v^2}}{\partial x_j}=\frac{1}{\rho}P_{\overline{v^2}}+R_{BP}+R_{NAT}-\omega\overline{v^2}-\frac{1}{\rho}D_{\overline{v^2}}+\frac{1}{\rho}\frac{\partial}{\partial x_j}\left[\left(\mu+\frac{\rho\alpha_T}{\sigma_k}\right)\frac{\partial\overline{v^2}}{\partial x_j}\right] \quad (2)$$

$$\frac{\partial\omega}{\partial t}+u_j\frac{\partial\omega}{\partial x_j}=\frac{1}{\rho}P_\omega+\left(\frac{C_{\omega R}}{f_W}-1\right)\frac{\omega}{\overline{v^2}}(R_{BP}+R_{NAT})-f_{NE}C_{\omega 2}\omega^2 f_W^2$$

$$+2\beta^*(1-F_1^*)\sigma_{\omega 2}\frac{1}{\omega}\frac{\partial k}{\partial x_j}\frac{\partial\omega}{\partial x_j}+\frac{1}{\rho}\frac{\partial}{\partial x_j}\left[\left(\mu+\frac{\rho\alpha_T}{\sigma_\omega}\right)\frac{\partial\omega}{\partial x_j}\right] \quad (3)$$

The SPF k $-\overline{v^2}-\omega$ model improves the nonequilibrium characteristics of turbulence based on the original k $-\overline{v^2}-\omega$ model. The modification term $f_{NE}$ in Eq. (3), which is multiplied by the destruction term of the ω equation, is used to produce more shear stress in the fully turbulent region. If $f_{NE}=1$, the model reverts to the original k $-\overline{v^2}-\omega$ model. In previous work [27,28], the shear layer region where $P_{\overline{v^2}}/\varepsilon>2.5$ is located by the switch function $\Gamma_{SSL}$, which means that the modification term $f_{NE}$ is turned off where $P_{\overline{v^2}}/\varepsilon$ is less than 2.5. The $Re_\Omega$ term in Eq. (4) is used to determine the magnitude of the modification, which indicates that the modification is enlarged in the large vorticity region away from the wall. The maximum value of $f_{NE}$ is chosen as 3.3 to remain unbounded.

$$f_{NE}=min(max(300Re_\Omega\Gamma_{SSL},1),3.3),\quad Re_\Omega=\frac{d^2\Omega}{\nu},\quad C_{\omega 2}=0.92 \quad (4)$$

$$\Gamma_{SSL}=\frac{1}{1+e^{-10(\frac{P_{\overline{v^2}}}{\varepsilon}-C_{SSL})}},\quad \frac{P_{\overline{v^2}}}{\varepsilon}=\frac{\mu_{T,s}S^2}{\rho\overline{v^2}\omega},\quad C_{SSL}=2.5 \quad (5)$$

The production terms are expressed as

$$P_k=\nu_T S^2,\quad P_{\overline{v^2}}=\nu_{T,s}S^2,\quad P_\omega=(C_{\omega 1}\frac{\omega}{\overline{v^2}}\nu_{T,s})S^2,\quad C_{\omega 1}=0.44 \quad (6)$$

The production-to-dissipation ratio $P_k/\varepsilon$ of the turbulent kinetic energy is a criterion for evaluating the nonequilibrium characteristics of the turbulence model. Within the context of the $k-\omega$ model framework, the production-to-dissipation ratio of the SPF model is obtained from

$$\frac{P_k}{\varepsilon} = \frac{\mu_T S^2}{\min(\rho\omega k, \rho\omega\overline{v^2})} \tag{7}$$

The nonequilibrium characteristics of the turbulence are reflected in the model constants. The constants of the original $k-\overline{v^2}-\omega$ model are calibrated based on several basic flows. One of the basic flows is a homogenous shear flow in which the gradient of the mean velocity is a constant. Note that $S = \partial u/\partial y$ dominates the velocity gradient, and the assumption can be written as

$$\frac{\partial k}{\partial x_i} = \frac{\partial \omega}{\partial x_i} = 0, \frac{\partial u}{\partial y} = const \tag{8}$$

With the above assumption, the three equations of the $k-\overline{v^2}-\omega$ model can be simplified to the following:

$$\frac{dk}{dt} = \frac{1}{\rho}P_k - \omega k \tag{9}$$

$$\frac{d\overline{v^2}}{dt} = \frac{1}{\rho}P_{\overline{v^2}} - \omega\overline{v^2} \tag{10}$$

$$\frac{d\omega}{dt} = \frac{1}{\rho}P_\omega - C_{\omega 2}\omega^2 \tag{11}$$

A solution of Eqs. (9)-(11) can be expressed as

$$\omega(t) = \omega_0, k(t) = k_0 e^{\lambda t} \tag{12}$$

$$\lambda = \frac{\Omega^2}{\omega} - \omega = \frac{P_k}{k} - \omega = \frac{\omega C_{\omega 2}}{C_{\omega 1}} - \omega = \omega\left(\frac{C_{\omega 2}}{C_{\omega 1}} - 1\right) \tag{13}$$

In addition, Eq. (9) can be written as

$$\frac{dk}{dt} = \frac{1}{\rho}P_k - \varepsilon \rightarrow \frac{\tau}{k}\frac{dk}{dt} = \frac{P_k}{\varepsilon} - 1, \quad \tau = \frac{k}{\varepsilon} = \frac{1}{\omega} \tag{14}$$

The quantity $\tau = \frac{k}{\varepsilon}$ has units of time and is called the turbulence timescale. An experiment showed that $\tau$ does not change appreciably, and the solution for $k$ is of the form of Eq. (15) [33]. Therefore, the kinetic energy grows exponentially over time in a homogenous shear flow, and the growth rate is related to $P_k/\varepsilon$. Comparing Eq. (12) and Eq. (15), one can see that the ratio of the coefficients $C_{\omega 2}/C_{\omega 1}$ is related to the production-to-dissipation ratio $P_k/\varepsilon$ and influences the transport behavior of the turbulent kinetic energy.

$$k(t) = k_0 e^{\frac{t}{\tau}(\frac{P_k}{\varepsilon}-1)} = k_0 e^{\omega(\frac{P_k}{\varepsilon}-1)t} \tag{15}$$

Measurements [24] indicate that the values of $P_k/\varepsilon$ are approximately $1.6 \pm 0.2$, where turbulence is in the equilibrium state. Using this, $C_{\omega 2}/C_{\omega 1}$ is usually calibrated as $1.4 < C_{\omega 2}/C_{\omega 1} < 1.8$ [24]. However, Rumsey et al. [29] and Li et al. [27,28] found that $P_k/\varepsilon$ can be greater than 1.8 in the nonequilibrium turbulence region. $P_k$ tends to be significantly larger than ε and often as much as 3-4 times greater in a separated shear layer. This leads to the hypothesis that increasing $C_{\omega 2}/C_{\omega 1}$ in the shear layer might be a correction for modeling nonequilibrium turbulence. The value of $C_{\omega 2}/C_{\omega 1}$ in the original k $- \overline{v^2} - \omega$ model is 2.09, which means that the original k $- \overline{v^2} - \omega$ model can model some of the nonequilibrium characteristics. In this paper, $f_{NE}$ is multiplied by the destruction term of ω to increase $C_{\omega 2}$ and $C_{\omega 2}/C_{\omega 1}$. The maximum value of $C_{\omega 2}/C_{\omega 1}$ in the present SPF k $- \overline{v^2} - \omega$ model is chosen to be 6.9 (only in the selected nonequilibrium region) to remain unbounded, and the modification is believed to be insensitive to this value.

The above discussion focuses on the relation between the constants $C_{\omega 1}$ and $C_{\omega 2}$ in a shear flow. In addition, these two constants are also relevant to the simulation of the boundary layer logarithmic region. Pop [33] illustrated that $C_{\omega 1}$ and $C_{\omega 2}$ must satisfy Eq. (16) for modeling the boundary layer logarithmic law in the k $-$ ε model. In the same way, the relation is also satisfied for the calibrations of $C_{\omega 1}$ and $C_{\omega 2}$ in this paper.

$$C_{\omega 1} = C_{\omega 2} - \frac{\kappa^2}{\sqrt{\beta^* \sigma_\omega}}, \beta^*=0.09, \sigma_{\omega 2}=1.17, \kappa=0.41 \tag{16}$$

The constants for the ω production and destruction terms of the SST model are calibrated for homogenous shear flow, and the model is an equilibrium turbulence model. The blending function of SST is used to switch the model between the k $-$ ω and k $-$ ε models in the near-wall region and bulk domain. In brief, the values of $C_{\omega 2}/C_{\omega 1}$ in the k $-$ ω and k $-$ ε models are 1.5 and 2.09, respectively. Compared with the SST, original k $- \overline{v^2} - \omega$, and SPF k $- \overline{v^2} - \omega$ models, we can see that the SPF k $- \overline{v^2} - \omega$ model has the strongest capacity to capture the nonequilibrium characteristics of turbulence and can automatically switch on the modification to increase the production-to-dissipation ratio in the separated shear layer. The original k $- \overline{v^2} - \omega$ model can also model a part of the nonequilibrium characteristics, while the SST model is almost in an equilibrium state in simulating a shear layer [28].

## III. Test Cases

**A. NLR7301 two-element airfoil**

The first high-lift configuration case is the NLR7301 two-element airfoil, which is a typical take-off configuration with a flap deflection of 20 degrees [7,8]. The model has been tested in an NLR Amsterdam 3 m×2 m low-speed wind tunnel and an NLR Northeast Polder 3 m×2.5 m low-speed wind tunnel. The overlap of the main wing and flap is 5.3%$c$, and the slot gap is 1.3%$c$. The Reynolds number based on the clean airfoil chord $c$ is 2.51 million, and the freestream Mach number is 0.185. The grid of this paper is generated based on the grid of ECARP [32], which is shown in Fig. 1. It is a C-type grid. The grid spacing of the first grid layer is $7.5 \times 10^{-6}$ m to ensure that $\Delta y^+$ is less than 1.0.

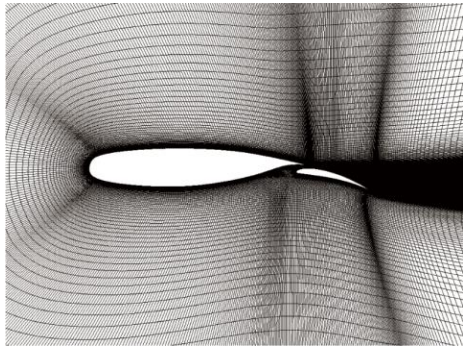

**Fig. 1 Computational grid of the NLR7301 configuration**

Fig. 2 shows comparisons of the turbulence models in terms of the aerodynamic coefficients. For $C_L - AOA$, the SST and SA models fail to predict the stall angle and the maximum lift coefficient. The $k - \omega$ model overpredicts the maximum lift coefficient. The predicted stall angle is postponed by approximately 2°. The original $k - \overline{v^2} - \omega$ model captures all lift characteristics well. The stall behavior is also effectively captured in the computations, although the rapid decrease in $C_L$ after the stall point is more moderate than with the experimental data. The results predicted by the SPF $k - \overline{v^2} - \omega$ model are almost the same as those of the original $k - \overline{v^2} - \omega$ model. The simulated drag coefficients are nearly identical for all models before the stall angle, and the predicted values are higher than the experimental data.

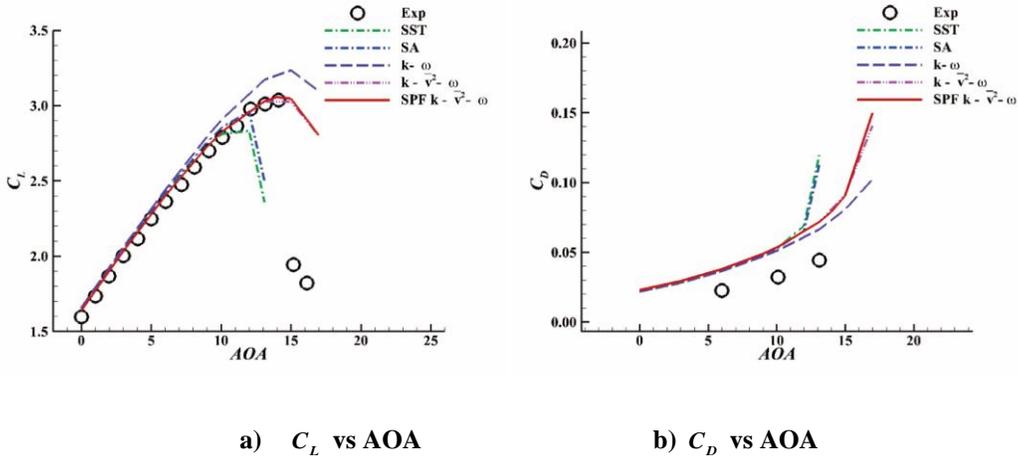

a) $C_L$ vs AOA  b) $C_D$ vs AOA

**Fig. 2 Comparisons of the different turbulence models in predicting the aerodynamic coefficients of the NLR7301 2-element airfoil (Ma=0.185, Re=$2.51 \times 10^6$)**

Comparing the pressure coefficient (Fig. 3) at AOA = 6°, one can observe that all models have nearly identical satisfying results. In contrast, the two-element airfoil is in a state of maximum lift at AOA = 13.1°. The original $k - \overline{v^2} - \omega$ and SPF $k - \overline{v^2} - \omega$ models predict nearly the same satisfying results, while the SA and SST models underestimate the suction peak. There is a persistent low-pressure suction platform at the trailing edge of the main wing, which means that the main wing experiences trailing edge separation. The $k - \omega$ model slightly overpredicts the height of the suction peak.

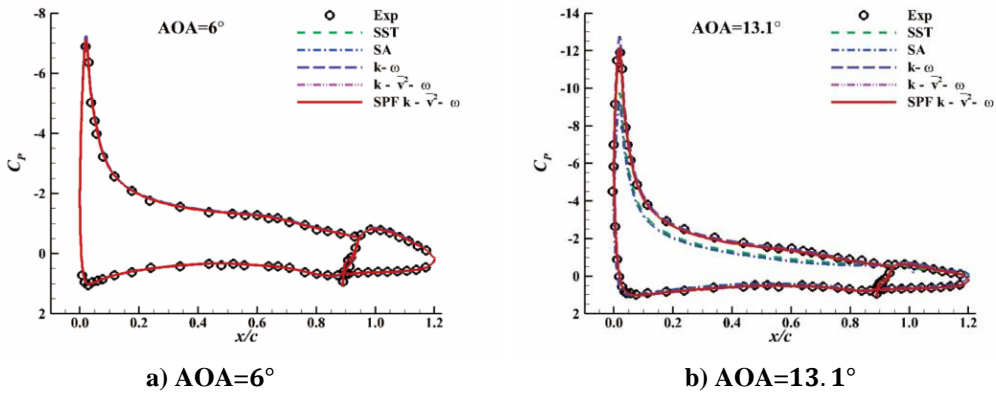

a) AOA=6°  b) AOA=$13.1°$

**Fig. 3. Comparisons of the turbulence models for the pressure distributions of the NLR7301 2-element airfoil**

The original $k - \overline{v^2} - \omega$ model accurately predicts the stall behavior for the NLR7301 airfoil. According to the relationship between $C_{\omega 2}/C_{\omega 1}$ and the nonequilibrium characteristics discussed in the previous chapter, the effects of the model constants are analyzed here. Changing the value of $C_{\omega 2}/C_{\omega 1}$ in the $k - \overline{v^2} - \omega$ model is equivalent to changing the modeling capability of the nonequilibrium turbulence of the model. We tested two values of $C_{\omega 2}/C_{\omega 1}$, 1.5 and 2.09, as shown in Fig. 4(a). According to Eq. (16), when $C_{\omega 2}/C_{\omega 1}$ =1.5, $C_{\omega 2}$ and $C_{\omega 1}$ are modified synchronously to 1.44 and 0.96, respectively. $C_{\omega 2}/C_{\omega 1}$ =1.5 is a typical model constant calibrated by an equilibrium turbulence assumption, and $C_{\omega 2}/C_{\omega 1}$ =2.09 is the baseline model constant used in the original $k - \overline{v^2} - \omega$ model. The predicted maximum lift coefficient and the stall angle of attack decrease with decreasing $C_{\omega 2}/C_{\omega 1}$. Comparing the pressure coefficient (Fig. 10(b)) at AOA = 13.1° , one can see that the suction peak is underestimated when $C_{\omega 2}/C_{\omega 1} = 1.5$. This phenomenon indicates that the ability to capture the nonequilibrium turbulence of the $k - \overline{v^2} - \omega$ model is very important for predicting the stall characteristics of the NLR7301 airfoil.

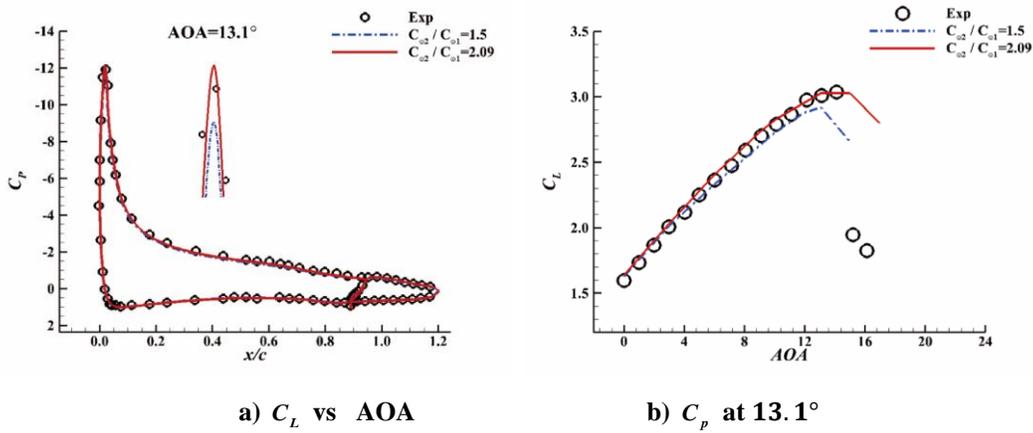

a) $C_L$ vs AOA  b) $C_p$ at $13.1°$

**Fig. 4 Comparisons for different values of $C_{\omega 2}/C_{\omega 1}$ in the $k - \overline{v^2} - \omega$ model in predicting the aerodynamic coefficients of the NLR7301 2-element airfoil**

The production-to-dissipation ratio $P_k/\varepsilon$ is obtained using the $k - \overline{v^2} - \omega$ model based on Eq. (15). Fig. 5 illustrates that the particular $P_k/\varepsilon$ predicted by the $k - \overline{v^2} - \omega$ model is significantly greater than 1.5 in the free shear layer. Such regions are the nonequilibrium turbulence regions. However, $P_k/\varepsilon$ is significantly reduced when $C_{\omega 2}/C_{\omega 1}$=1.5 because the turbulent kinetic energy decays rapidly along the streamwise location. This phenomenon illustrates that the $k - \overline{v^2} - \omega$ model loses the corresponding mechanism for capturing the nonequilibrium characteristics when $C_{\omega 2}/C_{\omega 1}$=1.5.

Fig. 6 compares the nondimensional velocities $U/U_{inf}$ under different $C_{\omega 2}/C_{\omega 1}$. The positions indicated by the red dashed lines in the figure signify the free shear layer formed by the large velocity gradient between the high-speed jet from the slot and the low-speed wake of the main element. The height of the main element wake increases significantly when $C_{\omega 2}/C_{\omega 1} = 1.5$, as shown in Fig. 6(b). Compared with that in the original $k - \overline{v^2} - \omega$ model, the enlarged wake height when $C_{\omega 2}/C_{\omega 1} = 1.5$ reduces the circulation of the airfoil, which significantly reduces the suction peak of the pressure distribution and leads to a decrease in the lift coefficient.

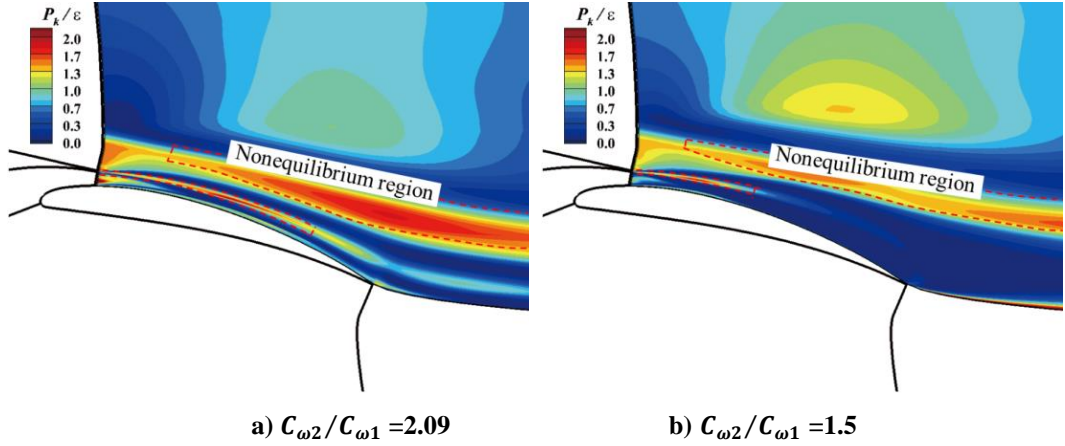

a) $C_{\omega 2}/C_{\omega 1}$ =2.09      b) $C_{\omega 2}/C_{\omega 1}$ =1.5

**Fig. 5 Comparisons of the $P_k/\varepsilon$ contours with different values of $C_{\omega 2}/C_{\omega 1}$ in the k $-\overline{v^2}-\omega$ model, AOA=10.1°**

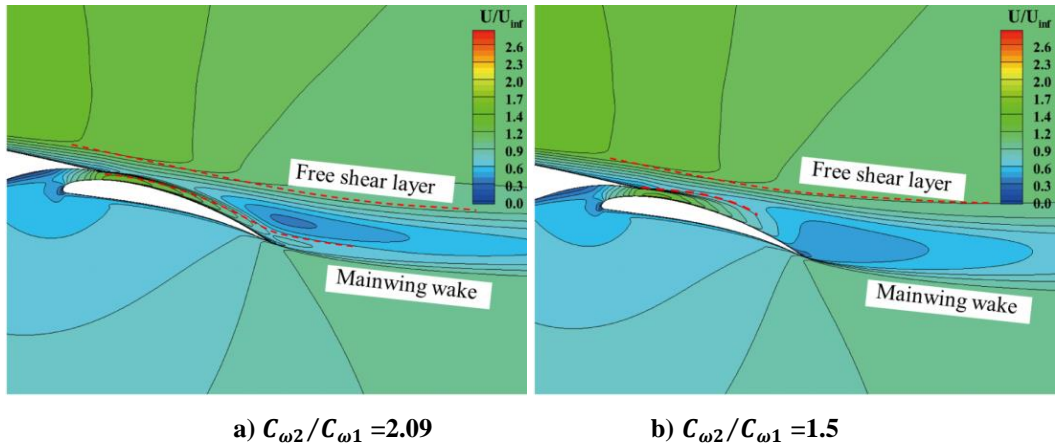

a) $C_{\omega 2}/C_{\omega 1}$ =2.09      b) $C_{\omega 2}/C_{\omega 1}$ =1.5

**Fig. 6 Comparisons of the $U/U_{inf}$ contours with different values of $C_{\omega 2}/C_{\omega 1}$ in the k $-\overline{v^2}-\omega$ model, AOA=10.1°**

**B. Omar four-element airfoil**

The second case is the Omar four-element airfoil, which was tested in a Boeing research wind tunnel in the 1970s [9]. Several configurations ranging from one element to five elements have been tested at Re = 2.83 million and M = 0.201. The double-slotted flap with a slat configuration discussed here is termed model C, as shown in Fig. 7(a). The deflection angles of the slat, the first flap, and the second flap are 50°, 0°, and 16.1°, respectively. Three meshes are applied to study the grid convergence, namely, a coarse mesh, medium mesh, and fine mesh, and the numbers of grids in these meshes are approximately $7 \times 10^4$, $1.4 \times 10^5$, and $2.5 \times 10^5$, respectively. A schematic diagram of the medium grid is shown in Fig. 7(b). The first grid layer $\Delta y^+$ values of the three grids are approximately 1.0, 0.6, and 0.4.

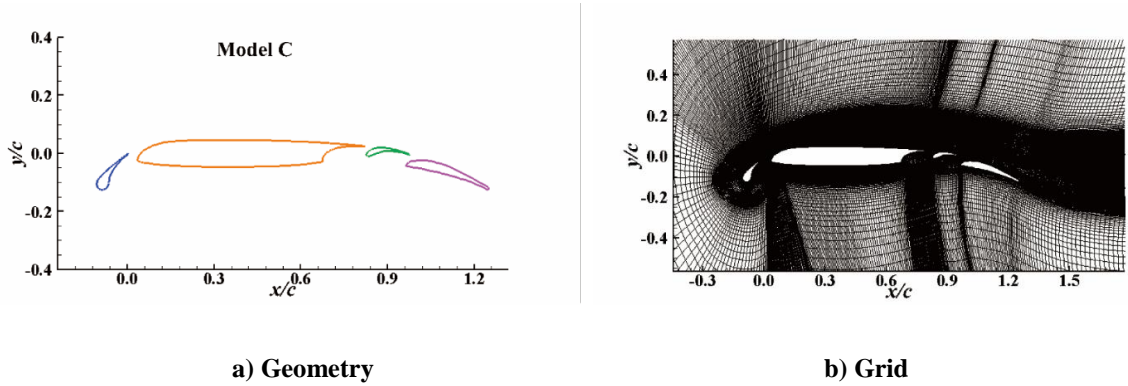

a) Geometry　　　　　　　　　　　　　　b) Grid

**Fig. 7 Geometry and grid of the Omar 4-element airfoil**

The $C_L - AOA$ curves of different grid densities for the four different turbulence models are presented in Fig. 8. The SA and SST models are sensitive to the grid size in the prediction of the maximum lift coefficient. The computed maximum lift coefficient increases gradually, and the stall angle of attack is delayed with increasing grid number. The $k - \omega$ model agrees well with the experimental data and has a good grid convergence performance except at the negative angles of attack of the coarse grid. The $k - \overline{v^2} - \omega$ model yields good agreement with the experiment through all AOAs. Additionally, the $k - \overline{v^2} - \omega$ model shows less sensitivity to the grid number. The coarse grid can also provide satisfying results.

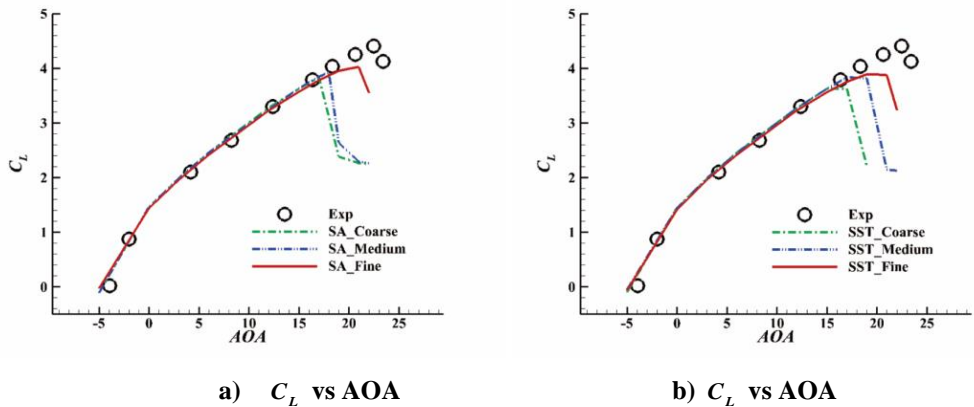

a)　$C_L$ vs AOA　　　　　　　　　　　　b)　$C_L$ vs AOA

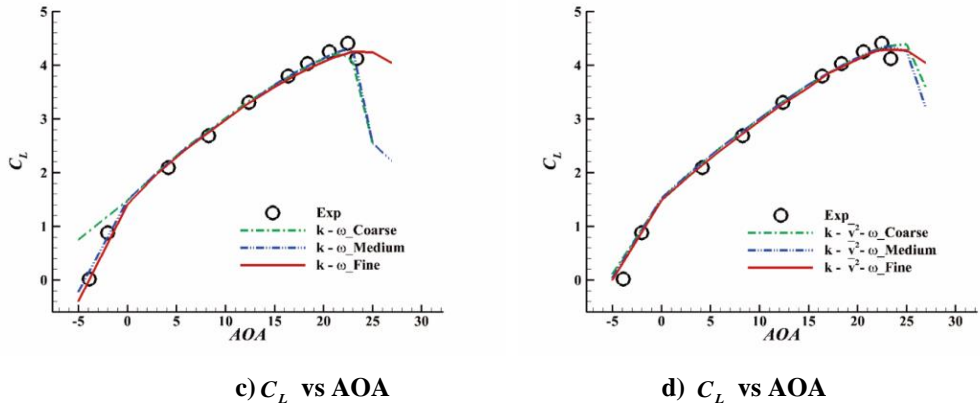

c) $C_L$ vs AOA  d) $C_L$ vs AOA

**Fig. 8 Grid convergence study for the Omar 4-element airfoil using four different turbulence models**

According to the analysis of the grid sensitivity, the fine grid is used to evaluate the accuracy of predicting the aerodynamic performance by different turbulence models. Fig. 9 shows comparisons of the RANS results of the five turbulence models. As shown in the figure, the SST and SA models are in good agreement with the measurements at the negative angles of attack and linear range but fail to predict the stall angle and the maximum lift coefficient. The $k - \omega$, original $k - \overline{v^2} - \omega$ and SPF $k - \overline{v^2} - \omega$ models have nearly identical satisfying results. The relative errors in the maximum lift coefficient for the SA, SST, $k - \omega$, original $k - \overline{v^2} - \omega$ and SPF $k - \overline{v^2} - \omega$ models are 9.03%, 12.65%, 4.28%, 5.19%, and 4.99%, respectively. The drag coefficients predicted by the five models have almost the same value except at the angles near the stall. When $C_L$ is lower than approximately 2.0, the drag is underpredicted compared with the experimental data, and the deviation of $C_D$ from the measurements decreases gradually. When the lift coefficient is higher than approximately 2.0, the drag is overpredicted, and the deviation of $C_D$ from the measurements increases gradually. All five models are nearly identical before the stall angle for predicting the moment coefficient. Comparing the pressure coefficient (Fig. 6(d)) at AOA = 22.13° where the four-element airfoil is in a state of maximum lift, the $k - \omega$, original $k - \overline{v^2} - \omega$ and SPF $k - \overline{v^2} - \omega$ models yield fine results, while the SA and SST models underestimate the suction peak, which is similar to the NLR7301 case.

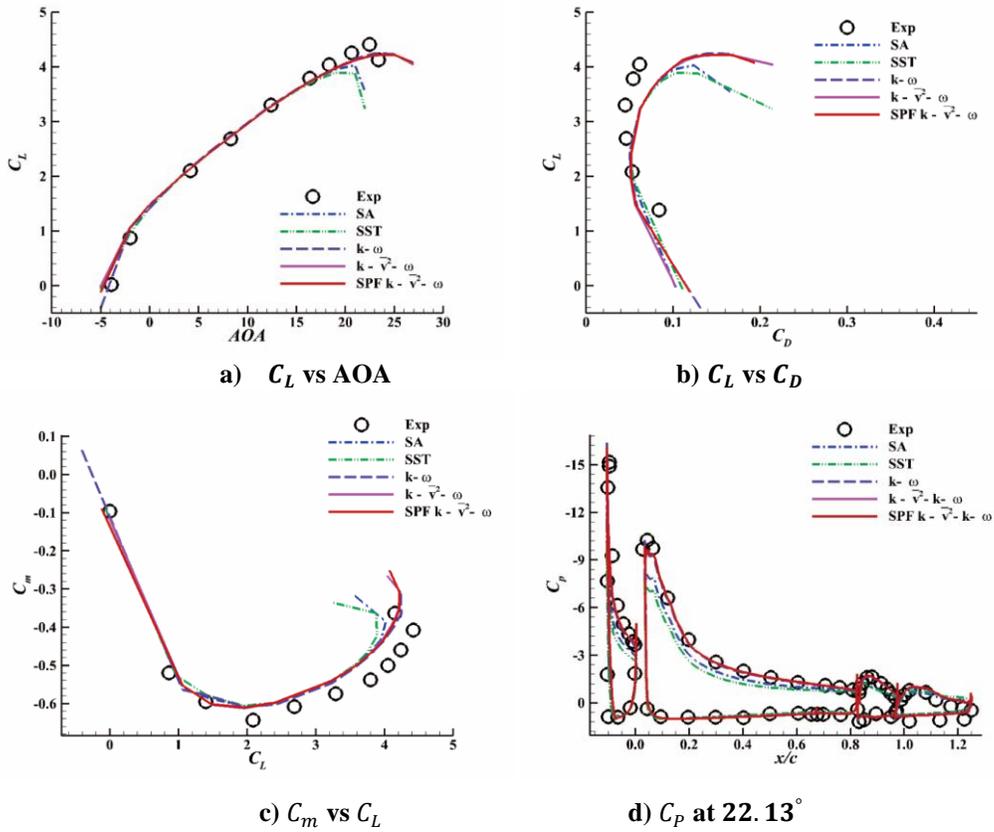

a) $C_L$ vs AOA

b) $C_L$ vs $C_D$

c) $C_m$ vs $C_L$

d) $C_P$ at $22.13°$

**Fig. 9 Comparisons of the different turbulence models in predicting the aerodynamic coefficients of the Omar 4-element airfoil**

The influence of the nonequilibrium characteristics of the $k - \overline{v^2} - \omega$ model on predicting high-lift flows is further studied for the Omar 4-element airfoil. The $C_L - AOA$ curves at values of $C_{\omega 2}/C_{\omega 1}$=1.5, 1.7, and 2.09 are presented in Fig. 10(a). According to Eq. (16), when $C_{\omega 2}/C_{\omega 1}$=1.5 and 1.7, $C_{\omega 2}$ and $C_{\omega 1}$ are modified synchronously to 1.44 and 0.96, respectively, and to 1.17 and 0.69, respectively, when $C_{\omega 2}/C_{\omega 1}$=1.7. The predicted maximum lift coefficient and the stall angle of attack decrease with decreasing $C_{\omega 2}/C_{\omega 1}$. Comparing the pressure coefficient at AOA = 22.13° (Fig. 10(b)), one can see that the suction peaks of all four elements are underestimated when $C_{\omega 2}/C_{\omega 1} = 1.5$ and 1.7.

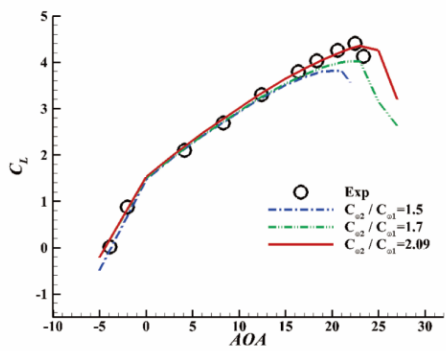 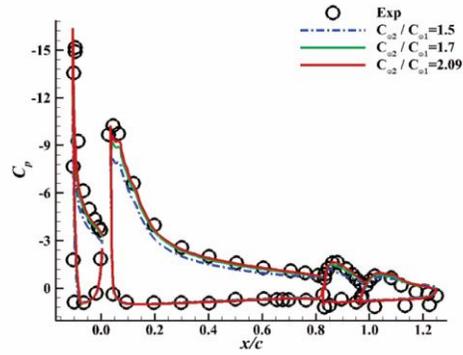

a) $C_L$ vs AOA  b) $C_P$ at $22.13°$

**Fig. 10 Comparisons of different values of $C_{\omega 2}/C_{\omega 1}$ in the $k - \overline{v^2} - \omega$ model in predicting the aerodynamic coefficients of the Omar 4-element airfoil**

Fig. 11 compares the nondimensional velocities $U/U_{inf}$ under the values of $C_{\omega 2}/C_{\omega 1} = 2.09$, 1.7, and 1.5. With a decrease in $C_{\omega 2}/C_{\omega 1}$, the wake height of the main element increases gradually and expands to the trailing edge of the main element, resulting in trailing edge separation. Because the nonequilibrium characteristics of the free shear layer cannot be captured, the height of the main wing wake increases with a decrease in the lift coefficient. This phenomenon is consistent with the results of the NLR7301 two-element airfoil and further demonstrates that capturing the nonequilibrium characteristics of turbulence in the free shear layer is a key factor in accurately predicting the stall behavior of the high-lift configuration.

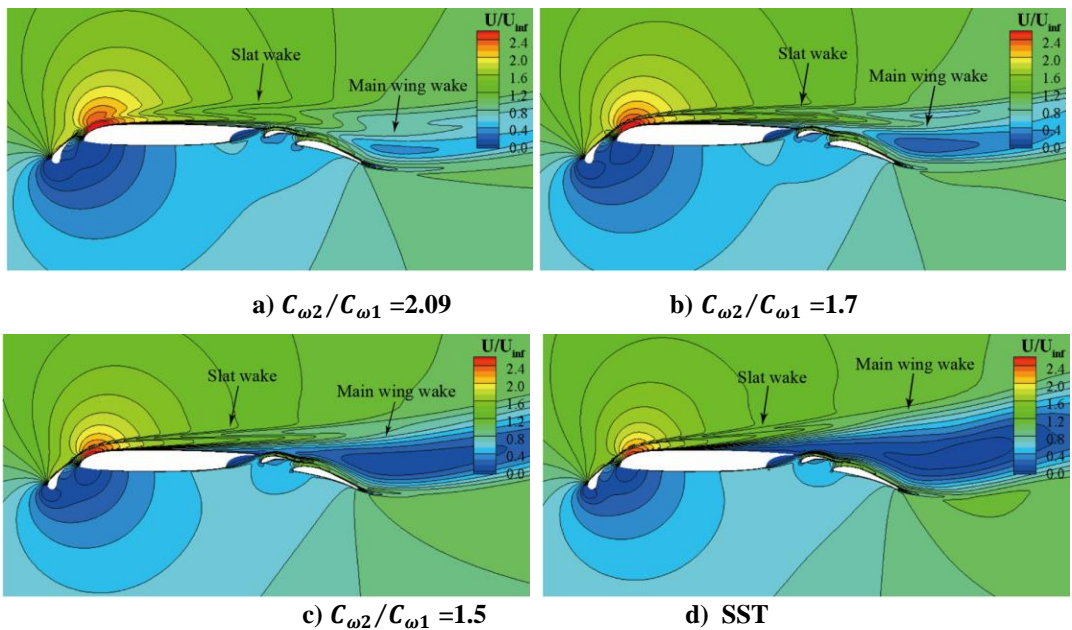

a) $C_{\omega 2}/C_{\omega 1} =2.09$   b) $C_{\omega 2}/C_{\omega 1} =1.7$

c) $C_{\omega 2}/C_{\omega 1} =1.5$   d) SST

**Fig. 11 Comparisons of the $U/U_{inf}$ contours with different values of $C_{\omega2}/C_{\omega1}$ in the k $-\overline{v^2}-\omega$ model, AOA=22.13°**

The angles of attack corresponding to the maximum lift coefficients for the NLR7301 and Omar airfoils predicted by the original k $-\overline{v^2}-\omega$ models are 13.1° and 22.13°. There are no obvious flow separation bubbles at these angles of attack, as shown in Fig. 6(a) and Fig. 11(a). This phenomenon is also in agreement with the finding of Rumsey [2], who indicated that many multielement airfoils exhibit no separated flow regions at the maximum lift coefficient. Because there is no obvious flow separation of the 2D multielement airfoils at high angles of attack near the stall, the separating shear layer fixed function in Eq. (5) is turned off. Consequently, the stall behavior predicted by the original k $-\overline{v^2}-\omega$ and SPF k $-\overline{v^2}-\omega$ models is nearly the same.

**C. JAXA Standard Model**

The third validation case is the high-lift configuration in the JAXA Standard Model (JSM), which is a representative half-span model of a realistic high-lift swept-wing regional jet airliner in landing configuration. The JSM was developed by JAXA and was chosen as the NASA Common Research Model for the 3rd High-lift Prediction Workshop [6]. The slat covers 90% of the leading edge. A configuration with a flap deflection angle of 30° and a slat deflection angle of 30° is selected. The overall dimensions and parameters are listed in Table 1.

**Table 1 Main dimensions of the JSM**

| Main dimension | | |
|---|---|---|
| Half span, s | [m] | 2.3 |
| Wing reference area, A/2 | [m$^2$] | 1.1233 |
| Reference chord, c$_{ref}$ | [m] | 0.5292 |
| Aspect ratio, $\Lambda$ | [-] | 9.42 |
| Taper ratio, $\lambda$ | [-] | 0.27 |
| Leading edge sweep, $\varphi_{LE}$ | [°] | 33 |
| Slat deflection angle, $\delta_s$ | [°] | 30 |
| Flap deflection angle, $\delta_f$ | [°] | 30 |

The configurations without (Case 2a) and with (Case 2c) nacelles/pylons were provided by HiLiftPW-3. A series of wind tunnel tests were performed on the JSM in the low-speed wind tunnel at JAXA, and high-quality test results were provided, including the aerodynamic forces, moments, and static pressure distributions at various span locations and an oil flow visualization. The test case presented in this paper corresponds to Case 2c of the workshop including nacelles/pylons, as shown in Fig. 12(a). Slat tracks and flap track fairings are also taken into account for the computations, as shown in Fig. 12(b). The Reynolds number based on the mean aerodynamic chord is 1.93 million, and the freestream Mach number is 0.172. The estimated tunnel turbulence intensity is 0.16%, and no transition trip is applied to the model. The test conditions are summarized in Table 2.

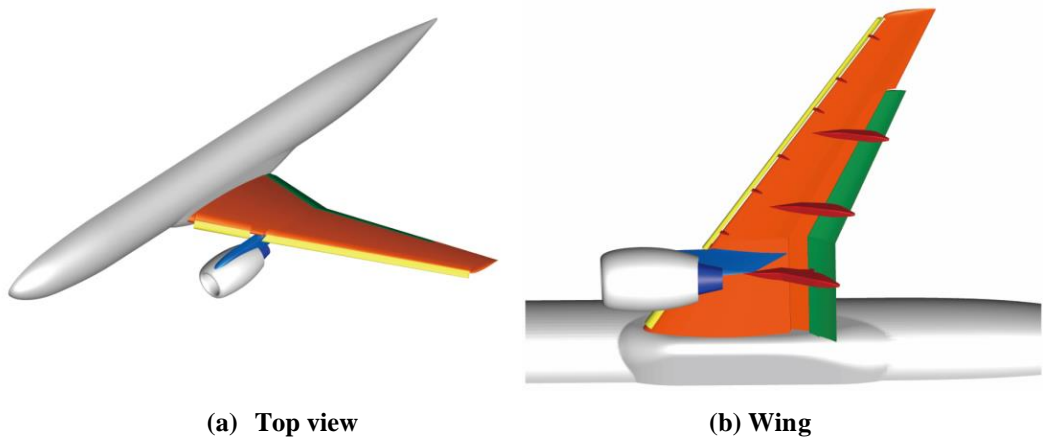

**(a) Top view**  **(b) Wing**

**Fig. 12 JSM wind tunnel model with nacelles/pylons**

**Table 2 Flow conditions for the JSM test**

| Flow conditions | | |
|---|---|---|
| Mach number | / | 0.172 |
| Reynolds number | / | 1.93 |
| Static pressure | [Pa] | 97000 |
| Static temperature | [K] | 306.55 |
| Turbulence intensity | [%] | 0.16 |

Several committee grids were created specifically for the workshop, and thus, a very large range of grid numbers is available. Due to the complex structure of the JSM, most of the committee grids are unstructured except the overlapping structured grid provided by NASA. There is no one-to-one structured grid. In this paper, we generated a multiblock one-to-one structured grid for the JSM. The surface grid is shown in Fig. 13.

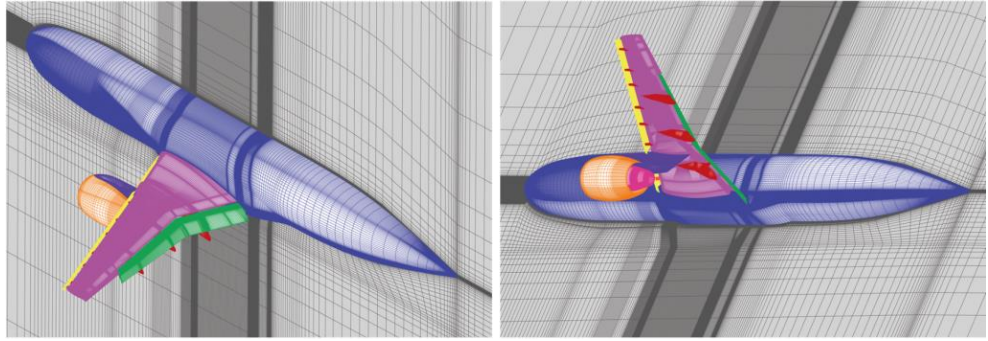

(a) Top view  (b) Bottom view

Fig. 13 Surface grid of the JSM

The grid convergence study is conducted with three sets of grids. The numbers of grids in the coarse grid, medium grid and fine grid are approximately 32 million, 56 million and 96 million, respectively. The first layer $\Delta y^+$ values for the three grids are approximately 1.0, 0.6, and 0.4. Fig. 14 presents the residual convergence histories of different grid densities calculated by the SPF $k - \overline{v^2} - \omega$ model at AOA=21.57 °. There are some fluctuations in the residual history of the coarse grid during the iterations. Compared with that of the coarse grid, the convergence histories of the medium grid and fine grid are better.

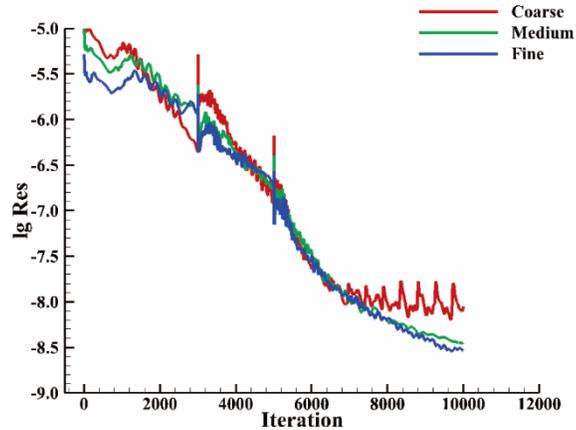

Fig. 14 Histories of the residuals in the coarse, medium, and fine grids, AOA=21.57 °

Fig. 15 compares the $C_L - AOA$, $C_L - C_D$, and $C_m - C_L$ curves obtained by the SPF $k - \overline{v^2} - \omega$ model with the experimental data. The maximum lift coefficient and stall angle are both underpredicted by the coarse grid. The lift curves are nearly the same for the medium and fine grids. The relative errors in the predicted maximum lift coefficient using the two grids are 3.61% and 1.44% compared with the experimental data. The coarse grid overpredicts the drag coefficient throughout the entire AOA range. The medium and fine grids yield better drag predictions at small lift coefficients but slightly overpredict the drag coefficient at large lift coefficients. The pitching moment curve is nearly the same for the medium and fine grids, and the variation tendency is in good agreement with the experimental data. Fig. 15(d) presents a comparison of the pressure coefficients at the 56% span station at AOA=18.58° predicted by the three grids. The medium and fine grids yield nearly the same results, which are in good agreement with the experimental data. However, the coarse grid underpredicts the suction peak of all three airfoil elements. In summary, the SPF model exhibits satisfactory grid convergence for the JSM high-lift configuration. The present paper validates the accuracy of the different turbulence models in predicting the stall behavior of the high-lift configuration. Therefore, the fine grid is used to perform CFD simulations to obtain accurate and reliable results in the following.

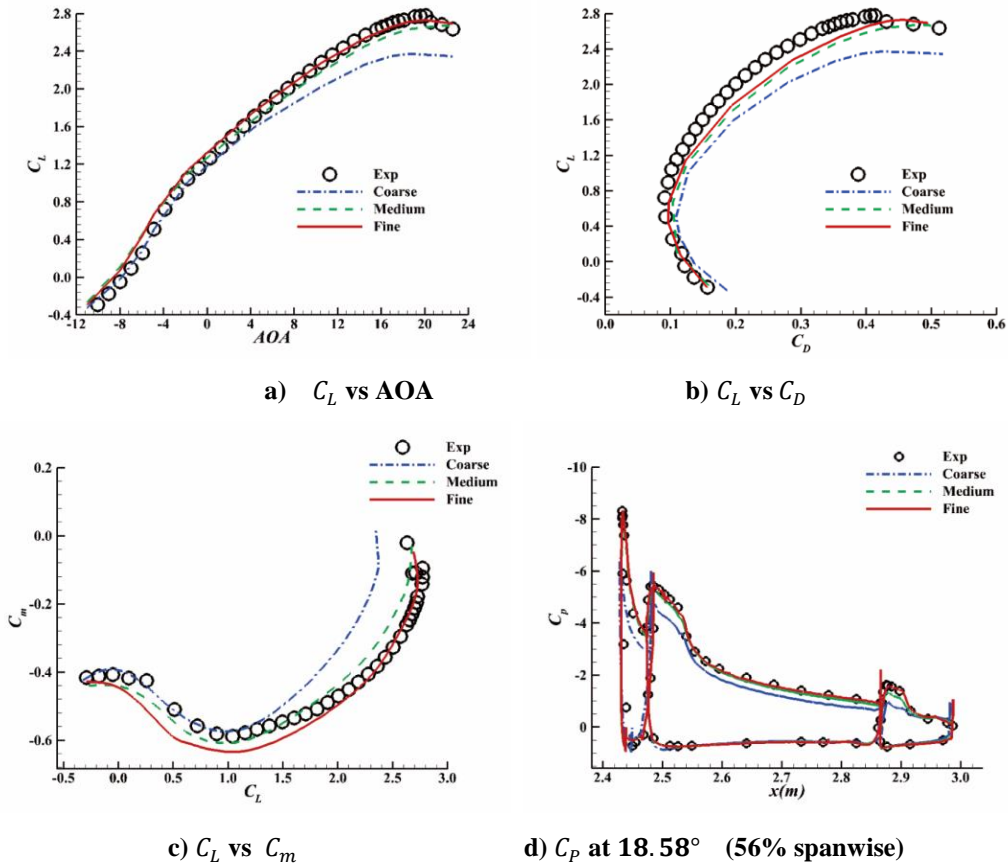

a) $C_L$ vs AOA  b) $C_L$ vs $C_D$

c) $C_L$ vs $C_m$  d) $C_P$ at $18.58°$ (56% spanwise)

**Fig. 15 Grid convergence study for the JSM using the SPF k $-\overline{v^2}-\omega$ model**

Fig. 16 shows comparisons of the turbulence models in terms of the aerodynamic coefficients. The SST and SA models fail to predict the stall angle of attack and the maximum lift coefficient. The original k $-\overline{v^2}-\omega$ model performs better than the SST and SA models in terms of predicting the stall angle; however, it slightly underestimates the maximum lift coefficient. The predicted stall angle of the original $k - \overline{v^2} - \omega$ model is 1.5° earlier than that of the experimental data and SPF model. The predicted maximum lift coefficients and stall angles of the different turbulence models are shown in Table 3. The relative errors in the maximum lift coefficient for the SST, SA, original $k - \overline{v^2} - \omega$ and SPF $k - \overline{v^2} - \omega$ models are 17.69%, 7.94%, 3.97%, and 1.44%, respectively. The drag coefficients predicted by the four models are nearly the same when $C_L$ is lower than 1.2 and in good agreement with the experimental data. When the lift coefficient is higher than 1.2, the drag is slightly overpredicted, and the deviation of the computed $C_D$ from the measurements increases gradually. The SPF $k - \overline{v^2} - \omega$ model yields fine results at a high angle of attack near the stall. For the pitching moment evaluation, the results predicted by the original $k - \overline{v^2} - \omega$ and SPF models are basically consistent with the experimental data, while the results predicted by the SST and SA models deviate from the experimental data at high angles of attack.

**Table 3 Maximum lift coefficients and stall AOAs of the different turbulence models**

|  | $C_{L,max}$ | $C_{L,max}$ point | Relative error |
|---|---|---|---|
| Experimental | 2.77 | 20.09 |  |
| SST | 2.28 | 14.54 | 17.69% |
| SA | 2.55 | 17.00 | 7.94% |
| Original $k - \overline{v^2} - \omega$ | 2.66 | 18.59 | 3.97% |
| SPF $k - \overline{v^2} - \omega$ | 2.73 | 20.09 | 1.44% |

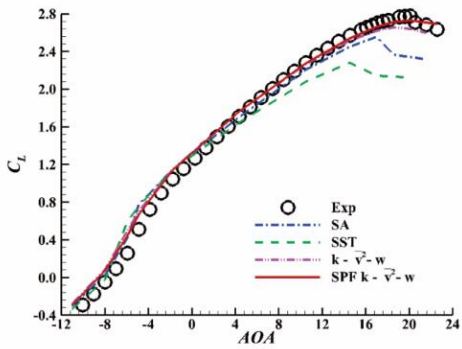
a) $C_L$ vs AOA

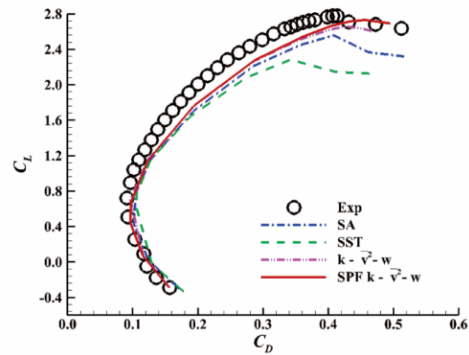
b) $C_L$ vs $C_D$

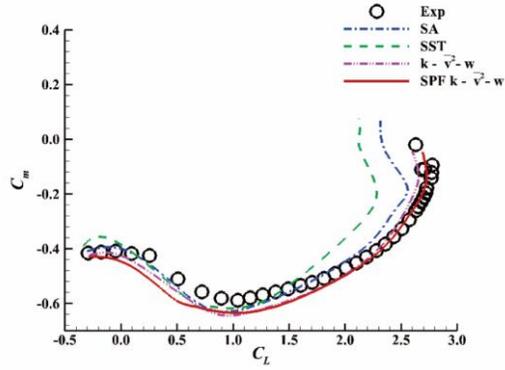

**c) $C_m$ vs $C_L$**

**Fig. 16 Comparisons of the different turbulence models in predicting the aerodynamic coefficients of the JSM high-lift configuration**

The pressure distributions at different span stations of the wing at AOA=18.58 ° are shown in Fig. 17. The SA and SST models underestimate the suction peak at all span stations, and the pressure coefficients demonstrate significant flow separation on the outer half of the wing along sections E-E and G-G. The original $k - \overline{v^2} - \omega$ model performs better than the SST and SA models in terms of the suction peak, and there is no separation on the outboard except the wing tip. The $k - \overline{v^2} - \omega$ and SPF $k - \overline{v^2} - \omega$ models predict higher suction peaks inboard and closer to the experimental value than the SA and SST models. A slight improvement is observed for the SPF modification, and the pressure distribution is closest to the experimental data.

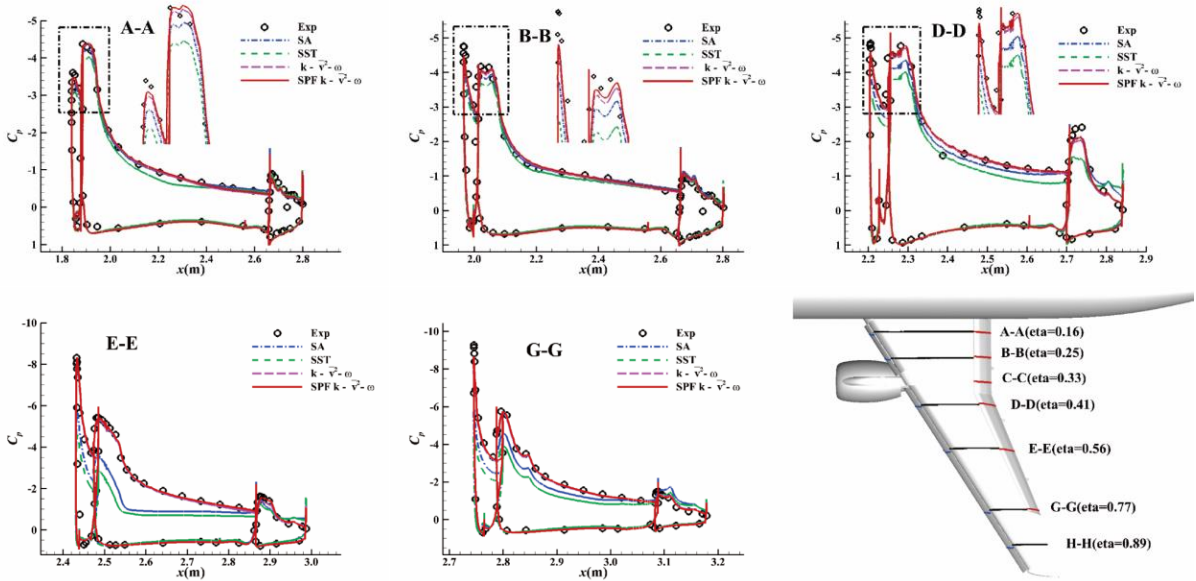

**Fig. 17 Pressure distribution results for the JSM, AOA = $18.58°$**

The post-stall surface visualizations predicted by the different turbulence models are shown in Fig. 18. The first and second images are obtained from oil flow measurements in the experiments at 18.58° and 21.57° angles of attack. The stall in the experiments starts from the wing root. The surface streamlines of the SA and SST models are plotted at 18.57° instead of 21.57° because of an early stall predicted by the two models. Strong flow separations are predicted by the two models on the outer wing. The original $k - \overline{v^2} - \omega$ model performs better than the SA and SST models, but trailing edge separation is predicted in the middle of the main element; thus, the maximum $C_L$ is slightly lower than the measured value. The stall behavior, as already seen in the integrated forces, is better predicted by the SPF $k - \overline{v^2} - \omega$ model. The separation on the wing outboard is similar to the oil flow image and is effectively predicted.

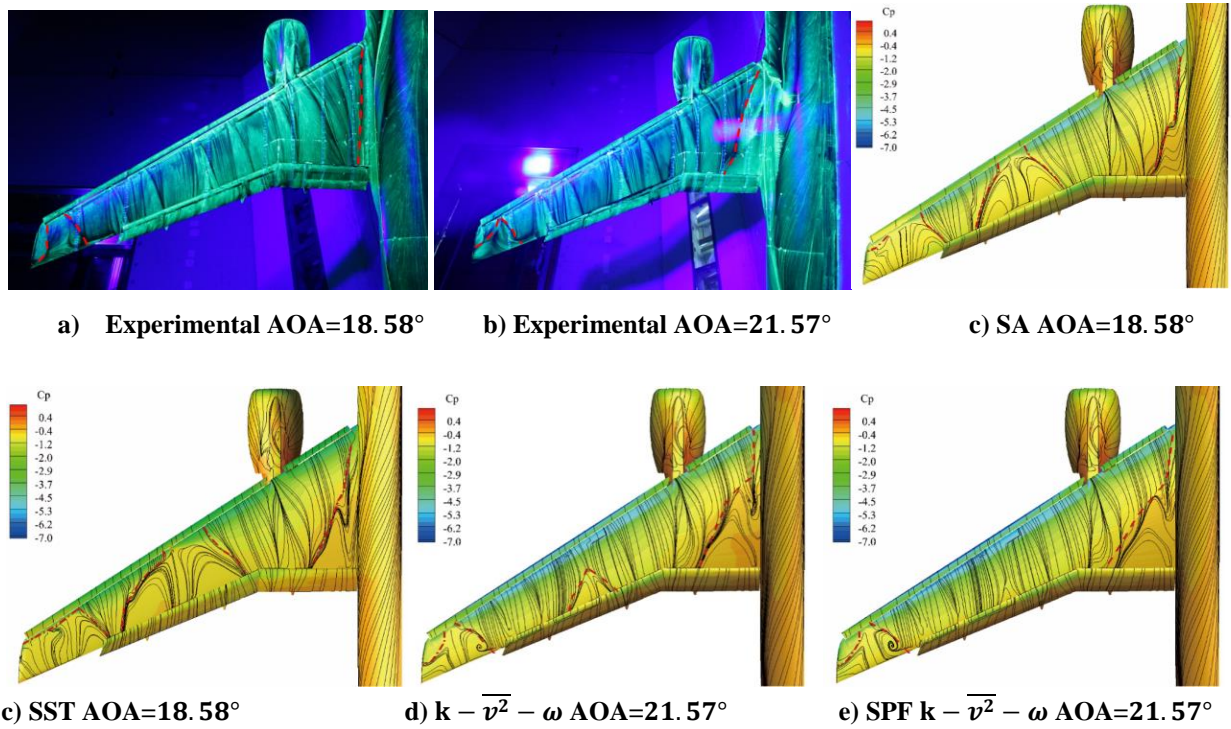

a) Experimental AOA=$18.58°$  b) Experimental AOA=$21.57°$  c) SA AOA=$18.58°$

c) SST AOA=$18.58°$  d) $k - \overline{v^2} - \omega$ AOA=$21.57°$  e) SPF $k - \overline{v^2} - \omega$ AOA=$21.57°$
**Fig. 18 Surface streamlines and oil flow pictures for JSM**

Fig. 19 presents the switch function $\Gamma_{SSL}$ contour at the 17.4% span station obtained using the SPF $k - \overline{v^2} - \omega$ model. When the switch function $\Gamma_{SSL}$ in Eq. (5) is larger than 1, the separating shear layer fixed modification is turned on. The main wing experiences trailing edge separation at the 17.4% span station at AOA=18.58°. The value of $\Gamma_{SSL}$ is greater than 1, and the modification is turned on at this separating shear layer. This phenomenon shows that the SPF modification can accurately locate the position of the separating shear layer.

Fig. 20 shows comparisons of the $P_k/\varepsilon$ contours at the 17.4% span station obtained using the $k - \overline{v^2} - \omega$ and SPF $k - \overline{v^2} - \omega$ models. This figure illustrates that the production-to-dissipation ratio predicted by the SPF model is significantly greater than 1.5 in the separating shear layer, which locates an area of the nonequilibrium turbulence region. The values of $P_k/\varepsilon$ obtained by the two models are basically the same in the free shear layer.

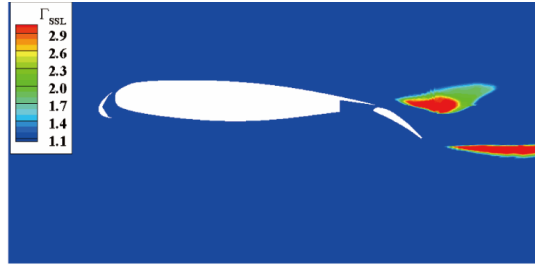

**Fig. 19 Switch function $\Gamma_{SSL}$ contour at the 17.4% span station obtained using the SPF $k - \overline{v^2} - \omega$ model, AOA=$18.58°$**

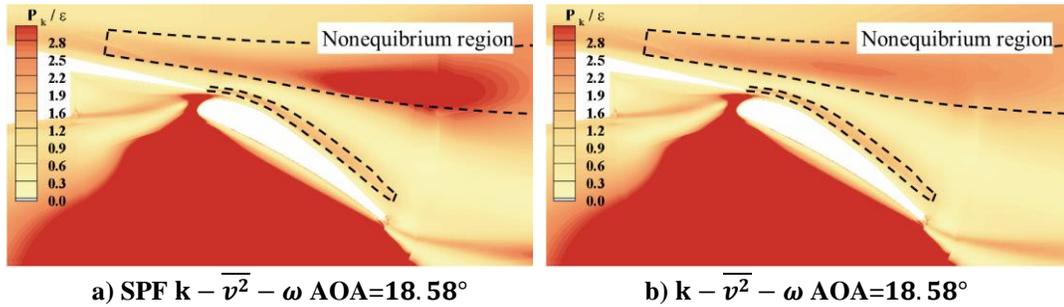

a) SPF $k - \overline{v^2} - \omega$ AOA=$18.58°$   b) $k - \overline{v^2} - \omega$ AOA=$18.58°$
**Fig. 20 Comparisons of the $P_k/\varepsilon$ contours at the 17.4% span station obtained using the different turbulence models**

Fig. 11 compares the nondimensional velocities $U/U_{inf}$ at the 17.4% span station obtained using the different turbulence models. The main wing experiences trailing edge separation at this span station. The main wing wake predicted by the original $k - \overline{v^2} - \omega$ model is slightly higher than that predicted by the SPF $k - \overline{v^2} - \omega$ model. Compared with that predicted by the SPF$k - \overline{v^2} - \omega$ model, the wake width overpredicted by the original $k - \overline{v^2} - \omega$ model reduces the circulation and decreases the lift coefficient. This is why the suction peak of the pressure distribution and maximum lift coefficient are underpredicted by the original $k - \overline{v^2} - \omega$ model. The low-speed wake predicted by the SST model is considerably larger than the wakes predicted by the two $k - \overline{v^2} - \omega$ models.

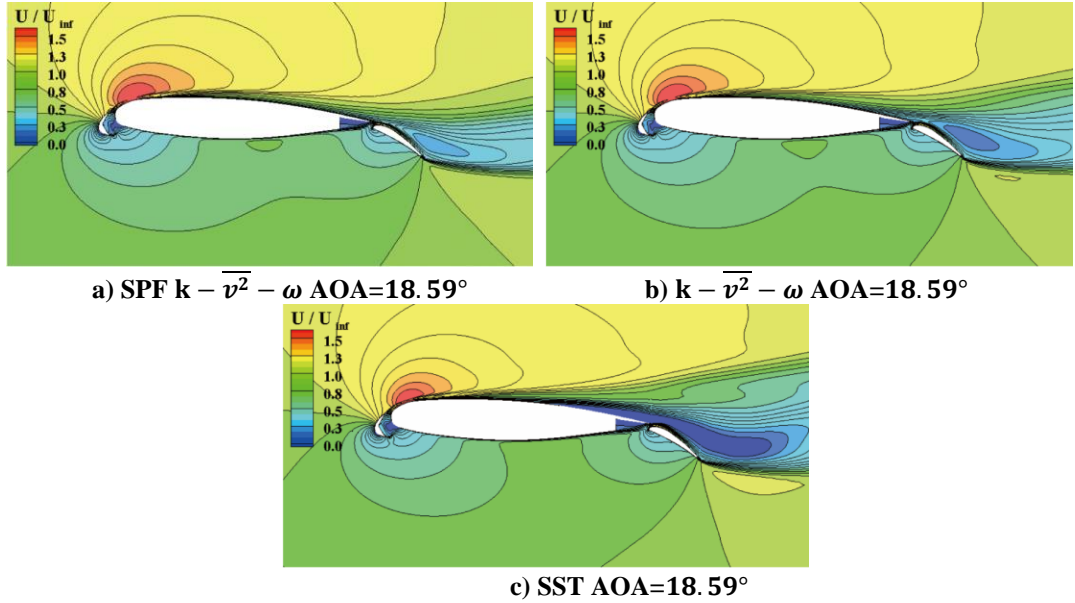

a) SPF k $-\overline{v^2}-\omega$ AOA=18.59°    b) k $-\overline{v^2}-\omega$ AOA=18.59°

c) SST AOA=18.59°

**Fig. 21 Comparisons of the $U/U_{inf}$ contours at the 17.4% span station obtained using the different turbulence models**

## IV. Conclusions

The complex flow of the high-lift configuration introduces difficulties to the CFD method, especially when using efficient turbulence models. In this paper, several RANS models are used to predict the stall performance of two-dimensional and three-dimensional high-lift devices. The selected turbulence models include the steady SA, SST, k $-\overline{v^2}-\omega$ and SPF k $-\overline{v^2}-\omega$ models. SPF is the abbreviation of the separating shear layer fixed, which is implemented on the k $-\overline{v^2}-\omega$ model. Such ad hoc fixes consider the nonequilibrium characteristics of turbulence in the shear layer. Through the analysis of the coefficients of the equations of the k $-\overline{v^2}-\omega$ models, it is found that the ratio of the coefficients $C_{\omega 2}/C_{\omega 1}$ represents the nonequilibrium characteristics of turbulence. The modified SPF k $-\overline{v^2}-\omega$ model fully considers the nonequilibrium characteristics of turbulence in the shear layer, while the SA and SST models are determined under equilibrium turbulence.

The multielement NLR7301 and Omar airfoils are numerically studied. These two cases show the typical flow fields of the two-dimensional high-lift configuration. The SA and SST models predict large separation bubbles near their predicted stall angles. In contrast, the k $-\overline{v^2}-\omega$ and SPF k $-\overline{v^2}-\omega$ models show no separation bubbles at these points, which is consistent with the observations in experiments. The complex three-dimensional model tested in this paper is the JAXA Standard Model. The SA and SST models predict strong flow separation on the outer wing.

The SPF k $-\overline{v^2}-\omega$ model accurately predicts the stall behavior of this configuration. The separation on the wing outboard is similar to the oil flow image and is effectively predicted.

For all the tested cases, the SA and SST models underpredict the stall angle and the maximum lift coefficient compared with the experimental results. These two models are sensitive to the grid density. The k $-\overline{v^2}-\omega$ and SPF k $-\overline{v^2}-\omega$ models yield satisfactory results on the multielement airfoils, while the SPF k $-\overline{v^2}-\omega$ model performs better on the complex full-configuration JSM. The relative errors in the predicted maximum lift coefficient are within 3% of the experimental data. Relative analysis and the results indicate that the nonequilibrium characteristics of turbulence are important in simulating the flow fields of the high-lift configuration.

## Acknowledgments

This work is supported by the National Natural Science Foundation of China under grant Nos. 11872230, 91852108 and 92052203.